\documentstyle[12pt]{article}
\begin{document}
\baselineskip 7mm
\title{ Generalized Aharonov-Bohm Effect, Homotopy Classes and Hausdorff
Dimension }
\author{H. Kr{\"o}ger \\ [2mm]
{\small\sl D{\'e}partement de Physique, Universit{\'e} Laval,
Qu{\'e}bec, Qu{\'e}bec, G1K 7P4, Canada } \\
{\small\sl Email: hkroger@phy.ulaval.ca } }
\date{June 1996 \\
Univesit\'e Laval preprint: LAVAL-PHY-6/96}
\maketitle
\begin{flushleft}
{\bf Abstract}
\end{flushleft}
We suggest as gedanken experiment a generalization of the Aharonov-Bohm
experiment, based on an array of solenoids. This experiment allows in principle
to measure the decomposition into homotopy classes of the quantum mechanical
propagator. This yields information on the geometry of the average path of
propagation and allows to determine its Hausdorff dimension.
\begin{flushleft}
PACS index: 03.65.Bz
\end{flushleft}
\setcounter{page}{0}

\newpage
\begin{flushleft}
{\bf 1. Introduction} \\
\end{flushleft}
In recent years, a number of very precise experiments have been
carried out in order to test the foundations of quantum mechanics.
One such fundamental property of quantum mechanics, which has never been
measured experimentally, is the zig-zagness of quantum mechanical paths of
propagation. Feynman and Hibbs \cite{kn:Feyn65} have noticed that
quantum mechanical paths are non-differentiable, statistically self-similar
curves.
Self-similarity is closely related to scaling, which plays an important r\^ole
in many areas of modern physics, e.g., in deep inelastics lepton-hadron
scattering,
Bjorken scaling in the parton model, quark distribution and splitting functions
in the Altarelli-Parisi equation. Mandelbrot \cite{kn:Mand83} has pointed out
that self-similarity is a characteristic feature of a fractal. Fractals are
characterized by a fractal dimension $d_{f}$ or a Hausdorff dimension $d_{H}$.
Abbot and Wise \cite{kn:Abbo81} have shown analytically that quantum mechanical
free motion yields paths being fractal curves with $d_{H}=2$.
Numerical simulations \cite{kn:Krog95} have shown
$d_{H} \neq 2$ to hold for velocity dependent potentials
like it occurs in Brueckner's \cite{kn:Brue55} theory of nuclear matter or via
dispersion relations for electrons propagating in solids
\cite{kn:Made81}.

\bigskip

Below we suggest a gedanken experiment which in principle allows to measure
the Hausdorff dimension of quantum mechanical paths.
In order to understand our choice of experimental set-up,
let us recall how to measure the Hausdorff (fractal) dimension $d_{H}$
of a fractal object in classical physics.
Mandelbrot \cite{kn:Mand83} considers as example the coastline of England.
One takes a yardstick, representing a straight line of a given length
$\Delta x$.
Let $\epsilon$ denote the ratio of the yardstick length $\Delta x$ to a fixed
unit length $l_{0}$.
Then one walks around the coastline, and measures the length of the coast with
the particular yardstick (starting a new step where the previous step leaves
off). The number of steps multiplied with the yardstick length
$\Delta x$ gives a value $L(\epsilon)$
for the coastal length. Then one repeates the same procedure with a smaller
yardstick of length $\Delta x'$, yielding the length $L(\epsilon')$. Eventually
one lets $\Delta x$ and hence $\epsilon$ go to zero.
One observes for a wide range of length scales $\epsilon$ that
the length of the British coast obeys a power law
\begin{equation}
L(\epsilon) = L_{0} \epsilon^{-\alpha}, \;\;\; (\epsilon \rightarrow 0).
\label{eq:criticalexp}
\end{equation}
This looks like the critical behavior of a macroscopic observable
at the critical point, thus $\alpha$ is called a critical exponent.
The Hausdorff dimension $d_{H}$ is defined by
\begin{equation}
d_{H} = \alpha + 1.
\label{eq:Hausdorffdim}
\end{equation}
As the example of the British coast line shows the determination of the
Hausdorff dimension of a curve requires to measure the length of curve with
many different length resolutions $\Delta x$. Then $d_{H}$ is defined only in
the limit $\Delta x \rightarrow 0$ via
the exponent of the power law.

\bigskip

Now suppose we want to do a corresponding experiment to study the geometry
of propagation of a massive particle in quantum mechanics. Position is an
observable in quantum mechanics. Thus one can monitor a particle being emitted
from a source at position $x_{in}$ at time $t_{in}$
to arrive at the detector at position $x_{fi}$ at time $t_{fi}$ by measuring
its position at intermediate times $t_{k}$ at regular intervals $\Delta t$.
This can be done experimentally as described in Ref.\cite{kn:Feyn65}. An
electron travelling from source to detector
passes by a number of screens with holes. In order to determine by which hole
the electron has passed the experimentator places a source of light behind each
screen emitting photons parallel to the screen.
Eventually, the photon collides with an electron having passed through a hole.
{}From detection of the scattered photon one can determine by which hole the
electron has passed. Thus one determines the length of path by joining the
source to the detector by the experimentally identified holes. This gives a
length $L(\Delta x)$ as a function of the resolution of length $\Delta x$,
being given by the size of holes and distance between the screens. In order to
extract the Hausdorff dimension one needs to send $\Delta x \rightarrow 0$,
i.e., decrease the size of holes and the distance between screens by increasing
the number of both.

\bigskip

This leads to the following problem: In order to localize with uncertainty
$\Delta x$ by which hole the electron has passed, one needs photon wave length
$\lambda < \Delta x$. Thus by the very measurement of position the electron
interacts with the photon and by collision has an uncertainty of momentum
$\Delta p \ge \frac{\hbar}{\Delta x}$. When going to the limit $\Delta x
\rightarrow 0$, the photon wavelength must go to zero and the uncertainty
of the electron momentum (in plane of screen) $\Delta p$ goes to infinity. Thus
the path becomes increasingly erratic. This can be interpreted by saying that
monitoring the path creates the fractal (erratic) path. Such an experiment does
not measure the geometry of propagation of a free quantum particle. This
dilemma occurs for every experiment which by any means measures the position of
the particle, i.e., monitors the path. One should mention also that Abbot and
Wise's \cite{kn:Abbo81} calculation of $d_{H}$ corresponds to localizing a wave
packet by position measurement thus monitoring the path, i.e., not to free
motion.
Strictly speaking, there is no analytical calculation of $d_{H}$ for
unmonitored paths.

\bigskip

It is the central theme of this letter to propose an alternative experiment
which avoids this dilemma and allows to determine the Hausdorff dimension of a
free particle without monitoring the path.
One can avoid to measure the position by using the concept of topology of
paths.
In the experiment described below one measures the interference pattern of the
cross section, and deduces the contributions of homotopy classes. In each
homotopy class the interaction with the vector potential of the magnetic field
is analytically known. Thus one can 'reconstruct' the non-interacting case.
Schulman \cite{kn:Schu71} has noted the
topological character of paths in the Aharonov-Bohm effect.
In the Aharonov-Bohm experiment an electron is scattered from
an (idealized) infinitely thin and long magnetic flux tube (solenoid).
We propose a generalized Aharonov-Bohm experiment consisting
of an array of such flux tubes.
An array of many flux tubes is needed because the spatial resolution $\Delta x$
is given in this experiment by the distance between neighbour solenoids. The
determination of $d_{H}$ requires $\Delta x \rightarrow 0$ hence many flux
tubes need to be placed between source and detector.
The array of flux tubes introduces a topology of paths. All paths (going from
$x_{in}$ to $x_{fi}$) can be classified by homotopy classes,
given by the number and sense of winding around each of the flux tubes of the
array. An electron propagating through the array of flux tubes
interacts with the vector potental of the static magnetic field. However,
for any path within a given homotopy class the corresponding electromagnetic
interaction is a constant, which is analytically computable. The problem is to
find out the relative weight of each homotopy class contributing
to the propagator. This is addressed by the experiment described below. \\

\newpage
\begin{flushleft}
{\bf 2. Aharonov-Bohm propagator} \\
\end{flushleft}
In order to understand the proposed experiment let us review the Aharonov-Bohm
experiment with a single flux tube and the corresponding calculation of the
quantum mechanical propagator.
For the case of the Aharonov-Bohm experiment with a single flux tube,
the corresponding homotopy classes are simple and the quantum mechanical
propagator in $2-D$ (plane perpendicular to flux) can be computed analytically
\cite{kn:Wilc90}.
We consider a charged particle (charge $q$) passing by (scattering from) the
solenoid (magnetic flux $\phi$). Classically, the Lorentz force is zero.
The gauge of the vector potential can be chosen
such that the vector potential takes the form
$A_{r}=0$, $A_{\theta}= \phi /2\pi r$.
The classical Hamiltonian in the presence of the vector potential is given by
\begin{equation}
H = \frac{1}{2\mu} \left( \vec{p} - \frac{q}{c} \vec{A} \right)^{2},
\label{eq:Hamiltonian}
\end{equation}
and the action is given by
\begin{equation}
S = \int dt \; \frac{\mu}{2} \dot{\vec{x}}^{2} +
\frac{q}{c} \dot{\vec{x}} \cdot \vec{A}(\vec{x},t).
\label{eq:action}
\end{equation}
Thus, when considering quantization by path integral,
there occurs an Aharonov-Bohm phase factor due to the vector potential
present in the action,
\begin{equation}
\exp \left[ \frac{i}{\hbar} \int_{0}^{T} dt \frac{q}{c} \dot{\vec{x}} \cdot
\vec{A} \right]
= \exp \left[ \frac{iq}{\hbar c} \int_{x_{in}}^{x_{fi}} d\vec{x} \cdot \vec{A}
\right]
= \exp[ i \alpha (\theta' -\theta + 2\pi n_{w})],
\label{eq:phasefactor}
\end{equation}
when the path winds $n_{w} = 0, \pm 1, \pm 2, \, \cdots $ times around the
solenoid, and $\alpha = q\phi/2 \pi \hbar c$. This factor depends only upon the
initial and final
azimutal angle $\theta$ and the number of windings, but otherwise it is
independent of the path. In other words, paths can be classified by their
winding number, they fall into homotopy classes.
The Aharonov-Bohm propagator ($2-D$) has been computed by Wilczek
\cite{kn:Wilc90}. It can be decomposed into contributions corresponding
to winding number $n_{w}$, given in spherical coordinates by
\begin{eqnarray}
&& K^{AB}_{n_{w}}(r',\theta';r,\theta)
\nonumber \\
&& = \int_{-\infty}^{+\infty} d \lambda
\exp[ i (\lambda + \alpha) (\theta' - \theta + 2 \pi n_{w}) ]
\frac{\mu}{2 \pi i \hbar T} \exp \left[ \frac{i \mu }{2 \hbar T}
(r'^{2} + r^{2} ) \right] \; I_{| \lambda |}\left( \frac{\mu r r'}{i \hbar T}
\right),
\label{eq:propagatorwind}
\end{eqnarray}
where $I_{\nu}(z)$ is the modified Bessel function. The free propagator
$K^{free}_{n_{w}}$ is given by $K^{AB}_{n_{w}}$ at $\alpha=0$.
Note that for each winding sector the Aharonov-Bohm propagator factorizes into
the Bohm-Aharonov phase and the free propagator,
\begin{equation}
K^{AB}_{n_{w}}(r',\theta';r,\theta)
= \exp[ i \alpha (\theta' - \theta + 2 \pi n_{w}) ]
K^{free}_{n_{w}}(r',\theta';r,\theta).
\label{eq:factorization}
\end{equation}
The total propagator (sum over all windings) is
\begin{equation}
K^{AB}(r',\theta';r,\theta) =
\sum_{m = -\infty}^{+\infty}
\exp[ i m (\theta' - \theta ) ]
\frac{\mu}{2 \pi i \hbar T} \exp \left[ \frac{i \mu }{2 \hbar T}
(r'^{2} + r^{2} ) \right] \; I_{| m -\alpha|}\left( \frac{\mu r r'}{i \hbar T}
\right).
\label{eq:propagator}
\end{equation}
When letting $r', r \rightarrow \infty$ the differential cross section
is obtained which has been firstly computed in a differerent way by Aharonov
and Bohm \cite{kn:Ahar59}.
Considering $r'=r$ to be large yields velocity $v=(r'+r)/T$,
momentum $p = \mu v$ and the de Broglie wave length
$\lambda = 2 \pi \hbar/p = \pi \hbar T/ \mu r$.

\bigskip

The original Aharonov-Bohm effect (one solenoid)
can be understood in terms of the semi-classical propagator \cite{kn:Feyn64}.
This holds when the distance $h$ between the solenoid and the classical path of
the electron (straight line between slit and detector) is large compared to the
de Broglie wave length $\lambda$, i.e., $h=\Delta x >> \lambda$ (classical
region).
The semi-classical propagator is defined as the free propagator times the
Aharonov-Bohm phase factor corresponding to the classical path.
However, in order to determine the Hausdorff dimension, $\Delta x$
needs to be sent to zero. thus the semi-classical case does not apply.
We have compared numerically the Aharonov-Bohm propagator
Eq.(\ref{eq:propagator}) with the semi-classical propagator. In Fig.[1] we show
the real part of the difference as a function of $\alpha$ and $h$. We kept the
following parameters fixed (given in dimensionless units): $x_{in}$, $x_{fi}$,
$L=2$(length of straight line between $x_{in}$ and $x_{fi}$), $T=10$, $\mu=1$,
$\hbar=1$.
We have chosen the cut-off $m_{max}=50$.
{}From convergence tests of the free propagator, we estimate that $m_{max}=20$
should be sufficient to guarantee stability in the sixth significant decimal
digit when $h \leq 10$.
The corresponding results for the imaginary part are similar.
This set of parameters corresponds to the de Broglie wave length
$\lambda = 10 \pi$ and the crossing of the quantum mechanical region to the
classical region occurs at $h=5$.
One observes that when the distance $h$ becomes large, the difference
between the Aharonov-Bohm propagator and the semi-classical propagator
tends to zero. However, one observes a marked difference for small distance $h$
($h \rightarrow 0$).
\\
\vspace{0.5cm}

\begin{flushleft}
{\bf 3. Generalized Aharonov-Bohm experiment} \\
{\bf (a) Set-up} \\
\end{flushleft}
The Aharonov-Bohm effect in the presence of one solenoid has been measured via
an electron interference experiment. Here I propose a generalization: An array
of $N_{S}$ solenoids is positioned in a regular array with next neighbour
distance
$\Delta x$ (see Fig.[2]). The array of solenoids is placed such that the
classical trajectory coming from slit A passes through this array,
while the classical trajectory coming from the slit B
does not pass through this array.
Like in the Aharonov-Bohm experiment one measures the interference pattern,
once when all solenoids are turned off and once when all solenoids are turned
on. Any change in the interference pattern is due to a
change of wave function which traverses the array of solenoids.
The values of $\phi_{i}$ (magnetic flux in solenoid $i$) are parameters
to be chosen by the experimentalist (see below).
The detector measures a squared modulus of the wave function
$I = \mid \psi(\vec{x},t) \mid^{2}$
and one observes an interference pattern. \\
\vspace{0.5cm }

\begin{flushleft}
{\bf (b) Homotopy classes} \\
\end{flushleft}
The quantum mechanical wave function can be expressed in terms of a path
integral (sum over paths),
\begin{equation}
\psi(\vec{x},t) =
\left. \int [dy] \; \exp[ \frac{i}{\hbar} S[\vec{y}] ]
\right|_{\vec{x},t;\vec{x_{0}},t}
= \sum_{C} \exp[ \frac{i}{\hbar} S[C] ],
\label{eq:wavefct}
\end{equation}
where the sum "over histories" goes over all paths $C$ starting from the source
at $\vec{x}_{0},t_{0}$ and going to the detector
at $\vec{x},t$ passing via either one of the two slits.
Because the action given by Eq.(\ref{eq:action}) has a free term and a magnetic
term the wave function can be factorized
\begin{equation}
\psi(\vec{x},t) =
\sum_{C} \; \exp[ \frac{i}{\hbar} S_{free}[C] ]
\; \exp[ \frac{iq}{\hbar c} \int_{C} d\vec{y} \cdot \vec{A}(\vec{y}) ]
= \sum_{C} \; K^{free}[C]
\; \exp[ \frac{iq}{\hbar c} \int_{C} d\vec{y} \cdot \vec{A}(\vec{y}) ].
\label{eq:splitwavefct}
\end{equation}
Quantum mechanical paths propagate forward in time, but can go forward and
backward in space. In $D \ge 2$ dimensions paths can form loops.
We have seen above that the Aharonov-Bohm propagator in a sector of fixed
winding number is given by the free propagator ($\alpha=0$) in this winding
sector times the Aharonov-Bohm phase factor. This Aharonov-Bohm phase factor is
the same for all those paths which can be mapped onto each other by stretching
and deformation without crossing the solenoid.
The winding number $n_{w}$ is a topological quantum number which characterizes
the paths. The full Aharonov-Bohm propagator is given by the sum over all
winding sectors.
All this carries over to the generalized Aharonov-Bohm experiment with
an array of $N_{S}$ solenoids. The full propagator decomposes into
homotopy classes. In each homotopy class the propagator factorizes into
the free propagator in this homotopy class and a generalized Aharonov-Bohm
phase factor, given in analogy to Eq.(\ref{eq:phasefactor}) by
\begin{equation}
\exp \left[ \frac{i q}{2 \pi \hbar c}
[ (\theta' -\theta) \phi_{tot}
+ 2\pi [ n_{1} \phi_{1} + \cdots n_{N_{S}} \phi_{N_{S}} ] \right],
\;\;\; \phi_{tot} = \phi_{1} + \cdots \phi_{N_{S}}.
\label{eq:genphasefactor}
\end{equation}
The topologically different (homotopy) classes are characterized by the winding
numbers $n_{1}, \cdots, n_{N_{S}}$, with $n_{i}=0,\pm 1,\pm 2, \cdots$.
Because Maxwell's theory is an Abelian gauge theory, homotopy classes do not
depend on the sequential order
of winding around individual solenoids. Equivalent paths with
the same winding, but different sequential order are shown in Fig.[3].

\bigskip

The decomposition property of the propagator for a fixed homotopy class
has the following implication being important for the experiment:
Changing the magnetic flux in the solenoid $\phi \rightarrow \phi'$ and hence
$\alpha \rightarrow \alpha'$, changes the Aharonov-Bohm
phase factor in each homotopy class and hence the total
Aharonov-Bohm propagator. But it {\it does not change}
the free propagator in each homotopy class.
Thus experimentally, one has a handle
to measure the free propagator corresponding to a given homotopy class.
We introduce a cut-off in the winding numbers $n_{i} < n_{cut-off}$.
This is based on the assumption that winding numbers beyond the cut-off
give contributions to the amplitude which are in the order of experimental
errors and hence can not be detected.
This cut-off makes the number of homotopy classes finite. Let us enumerate the
homotopy classes by $h=1,2, \cdots, N_{H}$.
The experimentalist chooses a set of fluxes of the solenoids:
$\phi^{(1)}_{i}, i=1,\cdots,N_{S}$ and measures the corresponding interference
pattern, say $I^{(1)}$.
Then the experimentalist chooses another set of fluxes of the solenoids,
$\phi^{(2)}_{i}, i=1,\cdots,N_{S}$, and measures again the interference
pattern,
$I^{(2)}$. This is repeated for $N_{F}$ differents sets of fluxes.
The information obtained is then sufficient to determine the
free propagators in the homotopy classes $h=1, \cdots, N_{H}$.
Substituting the phase factor, Eq.(\ref{eq:genphasefactor}), into the wave
function, Eq.(\ref{eq:splitwavefct}), yields the intensity
for $N_{F}$ different sets of fluxes,
\begin{equation}
I^{(f)} = \left|
\sum_{h}
K^{free}_{h}
\exp \left[ \frac{i q}{2 \pi \hbar c}
[ (\theta' -\theta) \phi_{tot}
+ 2\pi ( n_{1} \phi^{(f)}_{1} + \cdots n_{N_{S}} \phi^{(f)}_{N_{S}} ) ] \right]
\right|^{2}, \;\;\; f=1,\cdots,N_{F}
\label{eq:genintensity}
\end{equation}
Because a given set of fluxes and a given homotopy class $h$
determines the generalized Aharonov-Bohm phase factor,
and the free propagator in each homotopy class $K^{free}_{h}$
is independent from the fluxes,
this equation allows to determine the unknown coefficients
$K^{free}_{h}$ for $
h=1,\cdots,N_{H}$.
Because $K^{free}_{h}$ are complex numbers, and vector potentials $\vec{A}$
and fluxes $\phi$ and are real, we need at least twice as many sets of fluxes
as the number $N_{H}$ of homotopy classes considered, $ N_{F} > 2 N_{H} $. \\
\vspace{0.5cm}

\begin{flushleft}
{\bf (c) Length of paths and Hausdorff dimension} \\
\end{flushleft}
Suppose we have performed the above experiment and we know the free propagator
$K^{free}_{h}$ for homotopy classes $h=1, \cdots, N_{H}$.
{}From that we can construct the length of an average quantum mechanical path
in the following way.
Classically, one defines a length of a particle moving along a trajectory
(from $\vec{x}_{in}= \vec{x}(t_{in})$ to $\vec{x}_{fi}= \vec{x}(t_{fi})$)
by
\begin{equation}
L[x(t_{1}), x(t_{2}), \cdots, x(t_{N-1})] = \sum_{k=0}^{N-1} \mid
\vec{x}(t_{k+1}) - \vec{x}(t_{k}) \mid
\label{eq:classlength}
\end{equation}
and takes the limit $\Delta t \rightarrow 0$ in the end.
In quantum mechanics, position and length are observables.
In analogy to the classical mechanics
the definition of length  of trajectories in quantum mechanics also
involves the position (observable) at different times. In quantum mechanics
this requires to consider a transition amplitude from some initial state
$\mid \psi_{in} >$ at $t=t_{in}$ to some final state $\mid \psi_{fi} >$ at
$t=t_{fi}$. According to Feynman and Hibbs \cite{kn:Hibbs65} the transition
element for any function $F[x(t_{1}), x(t_{2}), \cdots, x(t_{N-1})]$
of position $x$ at different time steps $t_{1}, \cdots, t_{N-1}$ is given by
\begin{eqnarray}
< \hat{F} > &=&
\frac{ < \psi_{fi}(t_{fi}) \mid \hat{F}[x(t_{1}), \cdots , x(t_{N-1})] \mid
\psi_{in}(t_{in}) > }
{ < \psi_{fi}(t_{fi}) \mid \psi_{in}(t_{in}) > }
\nonumber \\
&=&
\frac{ \int [Dx(t)] dx_{fi} dx_{in} \; \psi_{fi}^{*}(x_{fi})
\; F[x(t_{1}), \cdots , x(t_{N-1})] \; \exp[ \frac{i}{\hbar} S ] \;
\psi_{in}(x_{in}) }
{ \int [Dx(t)] dx_{fi} dx_{in} \; \psi_{fi}^{*}(x_{fi})
\; \exp[ \frac{i}{\hbar} S ] \; \psi_{in}(x_{in}) }.
\label{eq:Feynaverage}
\end{eqnarray}
Feynman and Hibbs call this a weighted average. It can be interpreted as a sum
over all paths of the observable $F$ multiplied with the weight of the
exponential action.
Note that although this has an interpretation as path integral the matrix
element is a quantum mechanical expression which can be defined via the
Schr\"odinger equation. Substituting $F$ by the classical length,
Eq.(~\ref{eq:classlength}), and choosing position eigenstates as initial and
final states, one obtains $(x_{k} \equiv x(t_{k}))$
\begin{eqnarray}
< \hat{L}(\Delta t) > &=& < \sum_{k=0}^{N-1} | x_{k+1} - x_{k} | >
\nonumber \\
&=&
\frac{ \int d x_{1} \cdots d x_{N-1} \;
\sum_{k=0}^{N-1} | x_{k+1} - x_{k} | \; \exp[ \frac{i}{\hbar} S[x_{k}] ] }
{ \int d x_{1} \cdots d x_{N-1} \;
\exp[ \frac{i}{\hbar} S[x_{k}] ] }
\nonumber \\
&=&
\frac{ \sum_{C} L_{C} \exp[ \frac{i}{\hbar} S[C] ] }
     { \sum_{C} \exp[ \frac{i}{\hbar} S[C] ] }.
\label{eq:quantumlength}
\end{eqnarray}
The last equation is a short hand notation. Note that each curve $C$
corresponds to pieces of straight line joining positions at adjacent times,
i.e., $x(t_{k+1})$ with $x(t_{k})$ for $k=0,1,\cdots,N$.
Note, however, that this expression is not well defined in the limit
$\Delta t \rightarrow 0$. The average path is a fractal, hence its length
becomes infinite! This is an example, where an infinity occurs in the continuum
limit of non-relativistic quantum mechanics. We need to introduce a
regularization. A natural regularization of the transition element expressed
via the path integral is that given in Eq.(~\ref{eq:quantumlength}) where
$\Delta t$ is kept finite. However, such regularization is not suitable for the
proposed experiment because there is {\it no} measument taken at regular time
intervals $\Delta t$. On the other hand the experimentator has at his disposal
the spatial resolution $\Delta x$, i.e., the distance between neighbour
solenoids.
The resolution $\Delta x$ comes from an array of flux tubes. We have seen in
the previous sections that the path integral can be decomposed
into corresponding homotopy classes, counting the orientation and winding
number around each solenoid. Thus in analogy to the regularization of the path
integral via finite temporal resolution $\Delta t$ by
Eq.(~\ref{eq:quantumlength}), we define a regularization via finite spatial
resolution $\Delta x$ by
\begin{equation}
< \hat{L}(\Delta x) > = \frac{ \sum_{h=1}^{N_{H}} L(h) \exp[ \frac{i}{\hbar}
S[h] ] }
     { \sum_{h=1}^{N_{H}} \exp[ \frac{i}{\hbar} S[h] ] },
\label{eq:lengthregularization}
\end{equation}
where $h=1, \cdots, N_{H}$ denotes the homotopy classes, $N_{H}$ is the cut-off
determined from experiment, $\exp[ \frac{i}{\hbar} S[h] ] = K_{h}^{free}$
is the weight factor of the free action determined from the experiment and
$L(h)$ denotes the classical length of path in the homotopy class $h$.
In analogy to the regularization via $\Delta t$ by
Eq.(~\ref{eq:quantumlength}), where $L(C)$ is given by the classical length of
pieces of straight line, we define here $L(h)$ also by the classical length of
a curve being an element of homotopy class $h$. It starts at $x_{in}$ and
arrives at $x_{fi}$. It goes by pieces of straight lines always passing in the
middle of a pair of solenoids.
Such regularization does not distinguish paths on a scale smaller than $\Delta
x$. Thus the length $< \hat{L}(\Delta x) >$ is obtained by
taking $K^{free}_{h}$ for homotopy class $h$ from the experiment,
construct $L(h)$ for homotopy class $h$ from the array of flux tubes and
compute the sum according to Eq.(~\ref{eq:lengthregularization}).
This yields finally $< \hat{L}(\Delta x) >$ in absence of the vector potential,
i.e., corresponding to free propagation.
Finally, in order to extract the Hausdorff dimension $d_{H}$, one
has to measure the length
$<\hat{L}(\Delta x)>$ for many values of $\Delta x $,
look for a power law behavior when $\Delta x \rightarrow 0$ and determine the
critical exponent and thus $d_{H}$.
As a consequence of the fact that this experiment is not sensitive to the
zig-zagness parallel to the solenoids, we do not measure the length of the path
but only its projection onto the plane perpendicular to the solenoids, i.e., in
$D=2$ dimensions.
Nevertheless, the length as such is physically not so interesting
(it depends on $\Delta x$ anyway).
The physically important quantity is the critical exponent (Hausdorff
dimension)
which corresponds to taking the limit $\Delta x \rightarrow 0$.
But the latter should be the same in any number of space dimensions.

\bigskip

In summary, we have proposed a gedanken experiment how to measure the geometry
of propagation of a massive particle in quantum mechanics.
We have discussed the fundamental problem with experiments monitoring the path.
We suggest to avoid the problem by doing an experiment sensitive to the
topology of paths via a generalized Aharonov-Bohm experiment. This allows to
determine homotopy classes and the Hausdorff dimension. We call it a gedanken
experiment because we assume an idealized situation of infinitely thin flux
tubes.

\begin{flushleft}
{\bf Acknowledgement}
\end{flushleft}
The author is grateful for discussions with Prof. Franson, Johns Hopkins
University and Prof. Hasselbach, Universit\"at T\"ubingen.
The author acknowledges support by NSERC Canada.

\newpage
\begin{flushleft}
{\bf Figure Caption}
\end{flushleft}
\begin{description}
\item[{Fig.1}]
Absolut value of real part of difference between Aharonov-Bohm propagator
and semi-classical propagator.
Dependence on  distance $h$ and on $\alpha$.
Cut-off $m_{max}=50$.
\item[{Fig.2}]
Set-up of generalized Aharonov-Bohm experiment. There are $N_{S}$ solenoids
positioned in a regular array with distance $\Delta x$.
\item[{Fig.3}]
Example of two topologically equivalent paths.
\end{description}

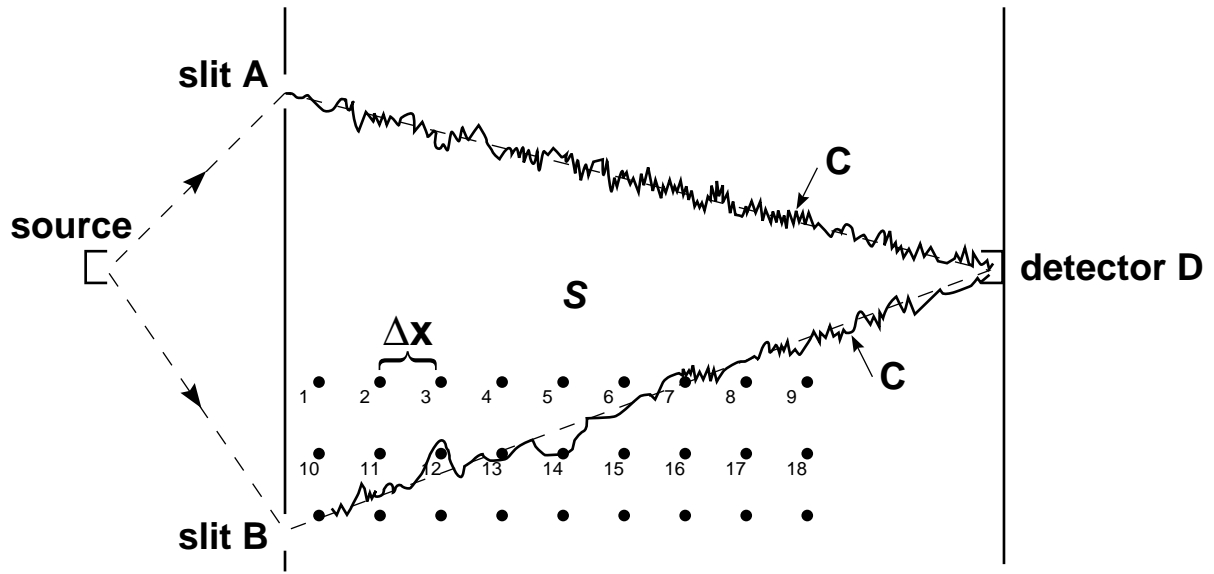
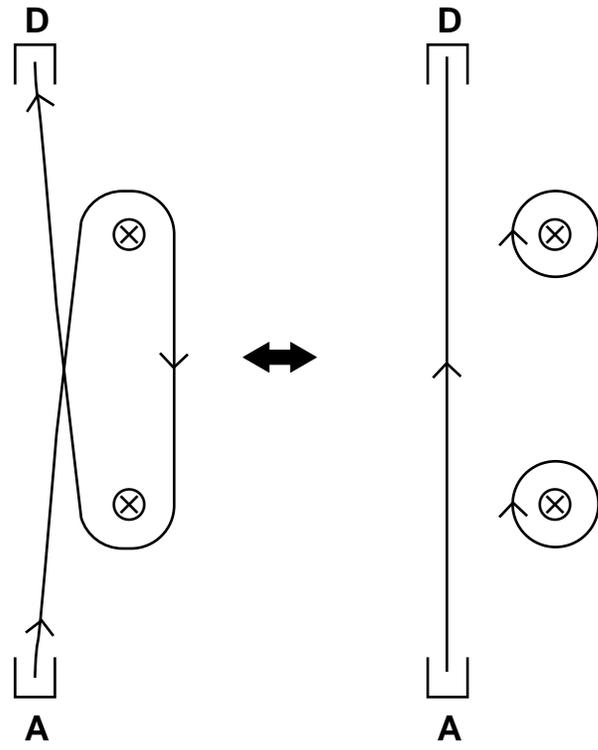
\end{document}